\documentclass[letterpaper, 10 pt, conference]{ieeeconf}  

\IEEEoverridecommandlockouts                              

\usepackage{tikz}
\usepackage{pgfmath}
\usetikzlibrary{chains,calc,quotes,intersections}
\usetikzlibrary{shapes,	shapes.geometric, shapes.symbols, shapes.arrows, shapes.multipart,	 shapes.callouts, shapes.misc}
\usetikzlibrary{automata, positioning}
\usetikzlibrary{arrows.meta}
\usetikzlibrary{matrix}
\usetikzlibrary{fit}
\usepackage{amsmath}
\usepackage{amssymb}
\usepackage{amsfonts}
\usepackage{graphicx}
\usepackage{glossaries}
\usepackage{epsfig}
\usepackage{epstopdf}
\usepackage{psfrag}
\usepackage{color}
\usepackage{ltablex}
\usepackage{ltxtable}
\usepackage{tabu} 
\usepackage{tabularx}  
\usepackage{flexisym}
\usepackage{float}
\usepackage{placeins}
\usepackage{textcomp} 
\let\labelindent\relax
\usepackage{enumitem}
\usepackage{subcaption}
\usepackage{multicol}
\usepackage[percent]{overpic}

\newacronym{icd}{ICDs}{implantable cardioverter defibrillators}
\newacronym{sut}{SUT}{system under test}
\newacronym{pmt}{PMT}{Pacemaker-Mediated Tachycardia}
\newacronym{as}{AS}{atrial sense}
\newacronym{vs}{VS}{ventricular sense}
\newacronym{ap}{AP}{atrial pacing}
\newacronym{vp}{VP}{ventricular pacing}
\newacronym{avi}{AVI}{Atrioventricular Interval}
\newacronym{savi}{sAVI}{AV Interval after sensing}
\newacronym{pavi}{pAVI}{AV Interval after pacing}
\newacronym{lri}{LRI}{Lower Rate Interval}
\newacronym{uri}{URI}{Upper Rate Interval}
\newacronym{esci}{ESCI}{Escape Interval}
\newacronym{vrp}{VRP}{Ventricular Refractory Period}
\newacronym{vb}{VB}{Ventricular Blanking Period}
\newacronym{arp}{ARP}{Atrial Refractory Period}
\newacronym{pvarp}{PVARP}{Postventricular Atrial Refractory Period}
\newacronym{tarp}{TARP}{Total Atrial Refractory Period}
\newacronym{pavb}{PAVB}{PostAtrial Ventricular Blanking}
\newacronym{aei}{AEI}{Atrial Escape Period}
\newacronym{bpm}{bpm}{beats per minutes}
\newacronym{mcps}{MCPS}{medical cyber-physical systems}
\newacronym{cps}{CPS}{cyber-physical systems}
\newacronym{erp}{ERP}{effective refractory period}
\newacronym{rrp}{RRP}{Relative refractory period}
\newacronym{av}{AV}{atrioventricular}
\newacronym{sa}{SA}{sinoatrial}
\newacronym{ams}{AMS}{automatic mode switching}
\newacronym{mvp}{MVP}{Managed Ventricular Pacing}
\newacronym{rvac}{RVAC}{Right ventricular automatic capture}
\newacronym{rsd}{RSD}{Rate smoothing down}
\newacronym{vsp}{VSP}{Ventricular Safety Pacing}
\newacronym{apc}{APC}{Atrial Premature Complex}
\newacronym{pvc}{PVC}{Ventricular Premature Complex}
\newacronym{pve}{PVE}{Ventricular Premature Events}
\newacronym{egm}{EGM}{electrogram}
\newacronym{elt}{ELT}{Endless Loop Tachycardia}
\newacronym{rnrvas}{RNRVAS}{Repetitive Nonreentrant VA synchrony}
\newacronym{apd}{APD}{Action Potential Duration}
\newacronym{cv}{CV}{Conduction Velocity}
\newacronym{hrv}{HRV}{Heart Rate Variability}
\newacronym{san}{SAN}{sinoatrial node}
\newacronym{avn}{AVN}{atrioventricular node}
\newacronym{ha}{HA}{hybrid automaton}
\newacronym{snrt}{SNRT}{sinus node recovery time}
\newacronym{fsm}{FSM}{finite-state machine}
\newacronym{hps}{HPS}{His-Purkinje system}
\newacronym{ra}{RA}{right atrium}
\newacronym{la}{LA}{left atrium}
\newacronym{rv}{RV}{right ventricle}
\newacronym{lv}{LV}{left ventricle}
\newacronym{ecg}{ECG}{electrocardiogram}
\newacronym{ta}{TA}{timed automaton}
\usepackage{array}

\tikzset{timing cycles/.style={
		rectangle,
		minimum size=5mm,
		very thick,
		draw=black,
		font=\itshape
	}}

\title{\LARGE \bf
An intracardiac electrogram model to bridge virtual hearts and implantable cardiac devices
}

\author{Weiwei Ai$^{1}$, Nitish Patel$^{1}$, Partha Roop$^{1}$, Avinash Malik$^{1}$, Nathan Allen$^{1}$ and Mark L. Trew$^{2}$
\thanks{$^{1}$Weiwei Ai, Nitish Patel, Partha Roop, Avinash Malik and Nathan Allen are with the Department of Electrical and Computer Engineering, University of Auckland, NZ
        {\tt\small wai484@aucklanduni.ac.nz}}%
\thanks{$^{2}$Mark L. Trew is with Auckland Bioengineering Institute, University of Auckland, NZ {\tt\small m.trew@auckland.ac.nz} 
}%
}

\begin{document}

\maketitle
\thispagestyle{empty}
\pagestyle{empty}

\begin{abstract}

Virtual heart models have been proposed to enhance the safety of implantable cardiac devices through closed loop validation. To communicate with a virtual heart, devices have been driven by cardiac signals at specific sites. As a result, only the action potentials of these sites are sensed. However, the real device implanted in the heart will sense a complex combination of near and far-field extracellular potential signals. Therefore many device functions, such as blanking periods and refractory periods, are designed to handle these unexpected signals. To represent these signals, we develop an intracardiac electrogram (IEGM) model as an interface between the virtual heart and the device. The model can capture not only the local excitation but also far-field signals and pacing afterpotentials. Moreover, the sensing controller can specify unipolar or bipolar \gls{egm} sensing configurations and introduce various oversensing and undersensing modes. The simulation results show that the model is able to reproduce clinically observed sensing problems, which significantly extends the capabilities of the virtual heart model in the context of device validation.

\end{abstract}

\section{INTRODUCTION}

With the growing use of implantable cardiac devices \cite{mond201111th}, it has become increasingly important to validate the devices under broader physiologically relevant conditions. This necessitates real-time virtual heart development to facilitate closed-loop validation. In closed-loop validation, virtual heart models provide physiologically relevant responses to the device \cite{jiang2012cyber,chen2014quantitative,DBLP:journals/corr/YipARMTAP16,ai2016requirements}. These models are usually based on \gls{ta} \cite{alur1994theory} or \gls{ha} \cite{henzinger1996theory}, which is amenable to formal analysis \cite{jiang2012cyber,chen2014quantitative} and real-time implementations \cite{allen2016modular,andalam2016hybrid}. This is desirable for the verification of safety-critical systems.

However, modeling the heart-device interface has received limited attention. In \cite{chen2014quantitative}, the heart model is connected with the device via specific computational nodes, which means that the device can only sense the immediate local signals. Synthetic \gls{egm} are generated by summing the distance-dependent Gaussian factor \cite{jiang2010real}. This can not reflect the real sensed signals, i.e. \gls{egm}, by the devices in patients, which include local activation, far-field signals and pacing artifacts \cite{Burri20171031}. 

The device senses these unexpected signals, referred to as \emph{oversensing}, which can cause inappropriate pacing inhibition, pacemaker tracking or mode switching \cite{Burri20171031,Swerdlow2017114}. For instance, an unintended mode switch is reported because of prolonged atrial activation \cite{almehairi2014unusual}. As a large proportion of the functionality of cardiac devices is designed to handle unexpected signals, we must be able to represent these behaviors in order to validate the functionality.

To the best of our knowledge, no existing work explicitly models far-field signals and pacing artifacts for the virtual heart \cite{jiang2012cyber,chen2014quantitative,DBLP:journals/corr/YipARMTAP16} in the context of closed-loop validation. Yip et.al \cite{DBLP:journals/corr/YipARMTAP16} use \gls{ha} to model an abstracted network of cardiac conduction system. The network is composed of cardiac cells connected by paths to capture characteristics of action potential propagation in tissue. This paper focuses on the approach to derive realistic \gls{egm} from the abstract heart model \cite{DBLP:journals/corr/YipARMTAP16}, but the methodology can also be applied to other work, like \cite{jiang2012cyber,chen2014quantitative,DBLP:journals/corr/YipARMTAP16}.

We develop an \gls{egm} model to capture local cardiac activities, stimuli afterpotentials and far-field signals. The model is designated for the virtual heart in the context of closed-loop validation. The same \gls{ha} formalism is employed, which enables us the amenability to formal analysis and real-time implementation. It expands the capability of validating timing cycles by introducing various oversensing and undersensing modes. The simulation results show that the model is able to reproduce sensing problems discussed in the literature like \cite{Burri20171031,Swerdlow2017114,almehairi2014unusual}.

\section{Materials and Methods}

\subsection{IEGM module}

\begin{figure}[!h]
	\centering
	
	\begin{tikzpicture}
	[->,>=stealth,auto,node distance=3.5cm, semithick,scale=0.7, transform shape,
	skip loop/.style={to path={-- ++(0,#1) -| (\tikztotarget)}},
	hv path/.style={to path={-| (\tikztotarget)}},
	vh path/.style={to path={|- (\tikztotarget)}}]
	\tikzstyle{every state}=[rectangle,rounded corners, fill=blue!10, align=center, 	minimum height = 2cm, text width=2.0cm, draw=none,text=black, draw,line width=0.3mm]
	
	\node[state]
	(Q0)  {\large $ Heart $};
	
	\node[state]
	(Q1) [node distance=4.5cm, right of=Q0] {\large $ IEGM$ };
	
	\node[state]
	(Q2) [node distance=4cm, right of=Q1] {\large $ Device$};
	
	\path[->] ($ (Q0.east)+(0,0) $) edge[double distance=1pt] node[align=center,shift={(0,0)}] {
		\footnotesize $Cell_i,Cell_j$\\
		\footnotesize $Relay_{i/j}$\\
		\footnotesize $Cell_{i\&j},Anni,...$
	} ( $ (Q1.west)+(0,0) $);
	
	\path[<-] ($ (Q0.east)+(0,-0.8) $) edge[double distance=1pt] node[align=center,shift={(0,0)}] {
		\footnotesize $AP,VP$  			
	} ( $ (Q1.west)+(0,-0.8) $);
	
	\path[->] ($ (Q1.east)+(0,0) $) edge[double distance=1pt] node[align=center,shift={(0,0)}] {
		\footnotesize $AEGM$ \\
		\footnotesize $ VEGM$ 			
	} ($ (Q2.west)+(0,0) $);
	
	\path[<-] ($ (Q1.east)+(0,-0.8) $) edge[double distance=1pt] node[align=center,shift={(0,0)}] {
		\footnotesize $AP,VP$  			
	} ($ (Q2.west)+(0,-0.8) $);
	
	\end{tikzpicture}		
	\caption{ IEGM module as the interface between the heart and the device.}
	\label{fig:interface}
\end{figure}
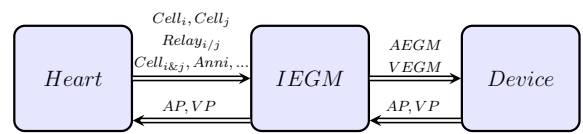

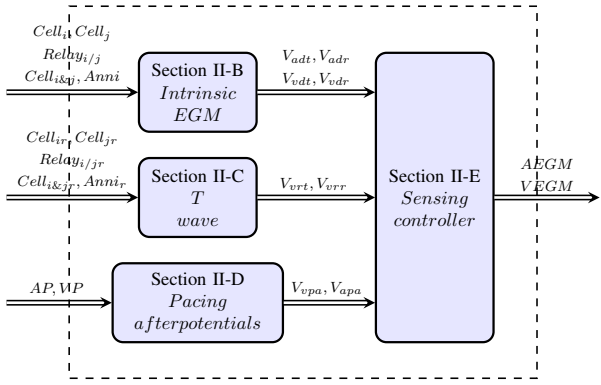
\begin{figure}[!h]
	\centering	
	\begin{tikzpicture}
	[->,>=stealth,auto,node distance=2cm, semithick,scale=0.7, transform shape,
	skip loop/.style={to path={-- ++(0,#1) -| (\tikztotarget)}},
	hv path/.style={to path={-| (\tikztotarget)}},
	vh path/.style={to path={|- (\tikztotarget)}}]
	\tikzstyle{every state}=[rectangle,rounded corners, fill=blue!10, align=center, minimum height = 1.5cm, text width=2.0cm, draw=none,text=black, draw,line width=0.3mm]
	
	\node[state]
	(Q0)  {Section \ref{sec:iegm}\\ $ Intrinsic$\\$ EGM $};
	
	\node[state]
	(Q1) [below of=Q0] {Section \ref{sec:t}\\$ T $\\$ wave$};
	
	\node[state,text width=3cm]
	(Q2) [below of=Q1] {Section \ref{sec:pacing}\\ $Pacing $\\ $afterpotentials$};
	
	\node[state,minimum height = 5.5cm]
	(Q3) [right of=Q1,node distance=4.5cm] {Section \ref{sec:sensing}\\$Sensing $\\ $controller$};	
	
	\path[->] ($ (Q0.west)+(-2.5,0) $) edge[double distance=1pt] node[align=center,shift={(0,0)}] {
		\footnotesize $Cell_i,Cell_j$\\
		\footnotesize $Relay_{i/j}$\\
		\footnotesize $Cell_{i\&j},Anni$  			
	} ( $ (Q0.west)+(0,0) $);
	
	\path[->] ($ (Q0.east)+(0,0) $) edge[double distance=1pt] node[align=center,shift={(0,0)}] {
		\footnotesize $V_{adt},V_{adr}$\\
		\footnotesize $V_{vdt},V_{vdr}$			
	} ( $ (Q3.west)+(0,2) $);	
	
	\path[->] ($ (Q1.west)+(-2.5,0) $) edge[double distance=1pt] node[align=center,shift={(0,0)}] {
		\footnotesize $Cell_{ir},Cell_{jr}$\\
		\footnotesize $Relay_{i/jr}$\\
		\footnotesize $Cell_{i\&jr},Anni_r$  			
	} ( $ (Q1.west)+(0,0) $);	
	
	\path[->] ($ (Q1.east)+(0,0) $) edge[double distance=1pt] node[align=center,shift={(0,0)}] {
		\footnotesize $V_{vrt},V_{vrr}$			
	} ( $ (Q3.west)+(0,0) $);	
	
	\path[->] ($ (Q2.west)+(-2,0) $) edge[double distance=1pt] node[align=center,shift={(0,0)}] {
		\footnotesize $AP,VP$  			
	} ( $ (Q2.west)+(0,0) $);
	
	\path[->] ($ (Q2.east)+(0,0) $) edge[double distance=1pt] node[align=center,shift={(0,0)}] {
		\footnotesize $V_{vpa},V_{apa}$			
	} ( $ (Q3.west)+(0,-2) $);	
	
	\path[->] ($ (Q3.east)+(0,0) $) edge[double distance=1pt] node[align=center,shift={(0,0)}] {
		\footnotesize $AEGM$ \\
		\footnotesize $ VEGM$ 			
	} ($ (Q3.east)+(2,0) $);
	
	\node[ draw, dashed,  inner xsep=1.9cm, inner ysep=1.6cm, shift={(0, 0.1)},
	label={[label distance=0cm]60:~}, 
	fit= (Q0)(Q1)(Q2)(Q3)] (BOX) {};
	
	\end{tikzpicture}		
	\caption{ Components of IEGM module.}
	\label{fig:IEGM}
\end{figure}

The proposed IEGM module acts as an interface between the heart and the device, as shown in Fig. \ref{fig:interface}, consisting of four components (Fig. \ref{fig:IEGM}). We extend the heart model \cite{DBLP:journals/corr/YipARMTAP16} with the events $Cell_i$, $Cell_j$ and so on to signify the propagation status, which are used to synchronize the \gls{egm} computation model detailed in Section \ref{sec:iegm}. Section \ref{sec:t} describes the T wave generated by the ventricular repolarization, pacing pulses $ AP$ and $VP $ from the device create pacing afterpotentials likely leading to crosstalk \cite{Burri20171031}, depicted in Section \ref{sec:pacing}. The sensing controller, Section \ref{sec:sensing}, is used to control the sensing configurations and introduce various sensing signals. The outputs atrial \gls{egm} ($ AEGM $) and ventricular \gls{egm} ($ VEGM $) go to the device sensing circuits.

\subsection{Intrinsic \gls{egm}}
\label{sec:iegm}
\begin{figure}[!h]
	\begin{center}
		\includegraphics[width=75mm]{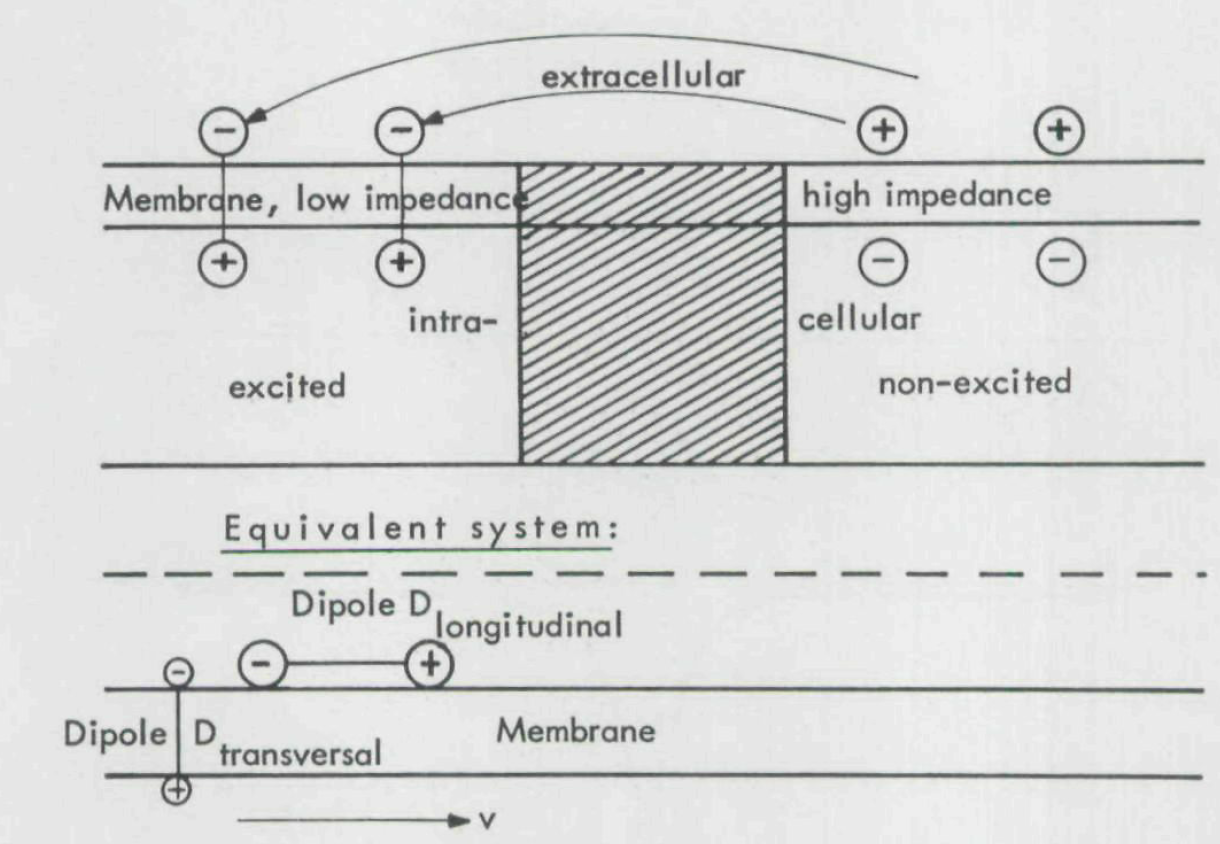}
		\caption{
			Dipole model \cite{irnich1985intracardiac}.
		}
		\label{fig:dipole}
	\end{center}
\end{figure}

The separation of electrical charges generated at the border between activated and resting myocardium creates moving electric dipoles \cite{wilson1933distribution,irnich1985intracardiac}, as shown in Fig. \ref{fig:dipole}. When the activation front moves nearby an electrode, the potential sensed by the electrode can be calculated by (\ref{eq1}) \cite{irnich1985intracardiac}, where $ C $ represents the dipole moment, $ r $ is the distance from the dipole to the electrode, and $ \varphi $ is the angle between the dipole and the distance vector. The amplitude decreases as the distance $ r $ increases. As a consequence, the local activation contributes much greater amplitude to the potential than far-field signals.

\begin{equation} \label{eq1}
V(r,\varphi) =C\cdot\frac{cos\varphi}{r^{2}}
\end{equation}

The heart model \cite{DBLP:journals/corr/YipARMTAP16} is composed of nodes and paths. The EGM computation module is connected with the path model. The event $Cell_i/Cell_j$ denotes that the action potential from $Cell_i/Cell_j$ starts traveling along the path. $ Relay_{i/j} $ means that the action potential reaches the other end of the path. $ Cell_{i\&j} $ represents both action potentials moving to each other and $ Anni $ indicates that they collide in the middle of the path.

\begin{figure} [!htp]
	\centering
	\begin{tikzpicture}
	[->,>=stealth,shorten >=1pt,auto,node distance=3.5cm, semithick,scale=0.65, transform shape,skip loop/.style={to path={-- ++(0,#1) -| (\tikztotarget)}},
	hv path/.style={to path={-| (\tikztotarget)}},
	vh path/.style={to path={|- (\tikztotarget)}}]
	\tikzstyle{every state}=[rectangle,rounded corners, align=left,
	minimum height = 1.7cm, text width=3.5cm, draw=none,text=black, draw,line width=0.3mm]		
	
	\node[state, text width=2.2cm]
	(Q0) [above of=Q0]{\large $ Idle $\\$ EGM_{ij}=0$};
	%
	\path[<-] (Q0.90) edge node[below, align=center, shift={(1.15,1)}] {} ++(0cm,1cm);
	\node[state,text width=3cm]
	(Q1) [node distance=4.5cm, left of=Q0] {\large $ EGM_{i} $\\ $\dot{t_{i}}=1 $ \\ $ V_i =C\cdot\frac{cos\varphi (t_i)}{r(t_i)^{2}} $\\$ EGM_{ij}=V_i$};
	
	\node[state,text width=3cm]
	(Q2) [node distance=4.5cm, right of=Q0] {\large $ EGM_{j} $\\ $\dot{t_{j}}=1 $ \\ $ V_j =C\cdot\frac{cos\varphi (t_j)}{r(t_j)^{2}} $\\$ EGM_{ij}=V_j$};	
	
	\node[state] 
	(Q3) [node distance=4cm, below of=Q0] {\large  $ EGM_{i\& j} $ \\ $ \dot{t_{ij}}=1 $\\$ V_i =C\cdot\frac{cos\varphi (t_i+t_{ij})}{r(t_i+t_{ij})^{2}} $\\$ V_j =C\cdot\frac{cos\varphi (t_j+t_{ij})}{r(t_j+t_{ij})^{2}} $\\$ EGM_{ij}=V_i+V_j $};		
	\path[->] (Q3) edge node[align=center,shift={(0,0)}] {
		\footnotesize $ Anni$? \\
	} (Q0);
	
	\path[->] ($ (Q0.west)+(0,0.3)$) edge node[align=center,shift={(0,0.8)}] {
		\footnotesize $Cell_i$? \\
		\footnotesize $ t_i\colon =0 $
	} ($ (Q1.east)+(0,0.3)$);
	
	\path[->] ($ (Q1.east)+(0,-0.5)$) edge node[align=center,shift={(0,0)}] {
		\footnotesize $ Relay_{i/j}$? 
	} ($ (Q0.west)+(0,-0.5)$);		
	
	\path[->] ($ (Q0.east)+(0,0.3)$) edge node[align=center,shift={(0,-0.1)}] {
		\footnotesize $Cell_j$? \\
		\footnotesize $ t_j\colon =0 $
	} ($ (Q2.west)+(0,0.3)$);
	
	\path[->] ($ (Q2.west)+(0,-0.5)$) edge node[align=center,shift={(0,0.5)}] {
		\footnotesize $ Relay_{i/j}$? 
	} ($ (Q0.east)+(0,-0.5)$);			
	
	\path[->] (Q1) edge[bend right] node[align=center,shift={(-1.2,-0.8)}] {
		\footnotesize $Cell_{ij}$? \\
		\footnotesize $ t_{ij} \colon=0 $
	} (Q3);
	
	\path[->] (Q2) edge[bend left] node[align=center,shift={(0,0)}] {	
		\footnotesize $Cell_{ij}$? \\
		\footnotesize $ t_{ij} \colon=0 $
	} (Q3);
	
	\end{tikzpicture}
	
	\caption{The model of the EGM computation.}
	\label{fig:egm}
\end{figure}
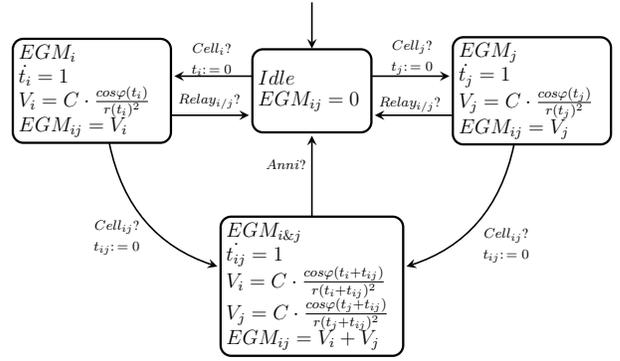

We assign each node with the coordinate $(x_i,y_i)$ and the conduction velocity $ v_{ci} $ along the path such that we can compute the instantaneous position of the moving action potential  $(x(t),y(t))$, $ cos\varphi(t) $, $ r(t) $ and then obtain the potential $ V $ at any specified electrode $(x_{p},y_{p})$ according to (\ref{eq1}) with the synchronization events.

Fig.\ref{fig:egm} shows the synchronizing control of the EGM computation with the $ Path_{ij} $. Given any electrode, the sensed EGM is the superposition of potentials generated by dipoles moving along all the paths within the heart model \cite{DBLP:journals/corr/YipARMTAP16}, including the local activation as well as far-field signals. 

In our model, we specify the coordinates of the tip and the ring electrodes in the right atrium and right ventricle. The sensed signals by these electrodes are denoted as $ V_{adt}, V_{adr}, V_{vdt}, V_{vdr} $ respectively.   

\subsection{T wave}
\label{sec:t}
The T wave reflects the ventricular repolarization \cite{Mirvis2015114}. In our model, we take the process of repolarization as a virtual dipole with advancing negative charges rather than the positive charges of the depolarization wave. The computation process is similar to Section \ref{sec:iegm}.

\subsection{Pacing afterpotential}
\label{sec:pacing}
A low-amplitude wave of opposite polarity is followed by a pacing pulse, referred to as \emph{afterpotential} \cite{Burri20171031}. The amplitude is directly related to both the amplitude and the duration of the pacing pulse. The afterpotential exponentially decreases and the decay characteristics are determined by the time constant formed by the product of the coupling capacitor C and the load (combination impedance of lead, electrode to tissue interface, and myocardium) \cite{sperzel2001reduction}, as shown in (\ref{eq2}).

\begin{equation} \label{eq2}
V(t)=V_{S}\cdot e^{\frac{-t}{RC}} 
\end{equation}

\begin{figure} [!htp]
	\centering
	
	\begin{tikzpicture}
	[->,>=stealth,shorten >=1pt,auto,node distance=3.5cm, semithick,scale=0.7, transform shape]
	\tikzstyle{every state}=[rectangle,rounded corners, align=left,
	minimum height = 1.7cm, text width=2.5cm, draw=none,text=black, draw,line width=0.3mm]	
	
	\node[state,text width=1.5cm]
	(Q0) []{\large $ Idle $ \\ $ V_{pa} =0$};
	%
	\path[<-] (Q0.90) edge node[below, align=center, shift={(1.15,1)}] {} ++(0cm,0.8cm);
	\node[state]
	(Q1) [node distance=6cm, right of=Q0] {\large $ PA$\\ $\dot{t}=1 $ \\ $ V_{pa} =V_S\cdot e^{\frac{-t}{RC}}$};	
	
	\path[->] (Q0.25) edge node[align=center,shift={(0,0)}] {
		\footnotesize $Stim_{end}$? \\
		\footnotesize $ t\colon =0 $
	} ($ (Q1.west)+(0,0.4) $);
	
	\path[->] (Q1.200) edge node[align=center,shift={(0,0)}] {
		\footnotesize $ Stim_{start}$? 
	} (Q0.330);		
	
	\end{tikzpicture}
	
	\caption{The model of the pacing afterpotential computation.}
	\label{fig:afterpotential}
\end{figure}
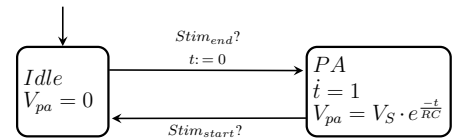

As shown in Fig. \ref{fig:afterpotential}, we can detect the start and the termination of the stimulation from the device, i.e., $AP$ or $ VP $, and employ (\ref{eq2}) to compute the afterpotential. The outputs $V_{vpa},V_{apa}$ denote atrial oversensing of ventricular pacing events (VA crosstalk) and ventricular oversensing of atrial pacing impulse (AV crosstalk).

\subsection{Sensing controller}
\label{sec:sensing}

Both oversensing and undersensing may trigger various algorithms of cardiac devices. These problems arise from many factors \cite{Burri20171031}, such as sensing threshold, configurations and so on. We develop a sensing controller to model these factors affecting EGM contents. In this module, AEGM and VEGM are the signals sensed by the atrial and ventricular leads, computed by (\ref{eq3}) and (\ref{eq4}) respectively. 

\begin{equation} \label{eq3}
AEGM=a\cdot(V_{adt}-b\cdot V_{adr} +c_{va}\cdot V_{vpa})
\end{equation}  

\begin{equation} \label{eq4}
\begin{split}
VEGM =&d\cdot(V_{vdt}-b\cdot V_{vdr}+ \\
&e\cdot (V_{vrt}-b\cdot V_{vrr}) +c_{av}\cdot V_{apa})
\end{split}
\end{equation}

The variables $ a,b,c_{va},c_{av},d,e $ are used to determine which signals are present to the device, including desired local activations and far-field signals $ V_{adt}, V_{adr}, V_{vdt}, V_{vdr} $, T wave $ V_{vrt}, V_{vrr} $ and afterpotentials $V_{vpa},V_{apa}$. These variables can be changed over time according to the validation requirements.

\section{Results and discussion}
\label{sec:results}

\begin{figure*}[]
	\begin{center}
		
		\begin{overpic}[height=2.4in]{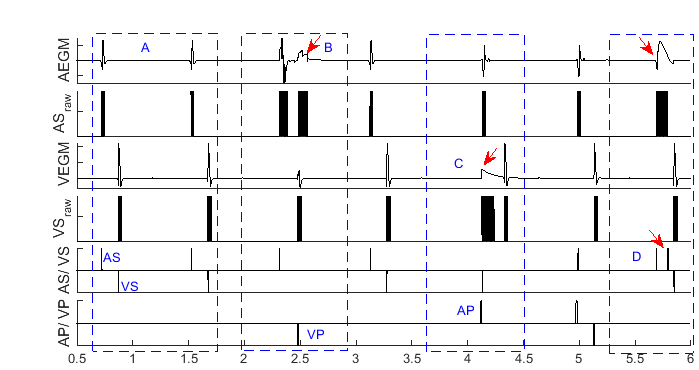}
			\put(50,-2){Time (s)} 
		\end{overpic}	
		\caption{
			Simulation EGMs. Segment $ A $ demonstrates desired AEGM/VEGM. $ B $ includes far-field R wave indicated by the red arrow.  AV crosstalk is present in $ C $ due to the atrial pacing pulse $ AP $. In $ D $, the prolonged P-wave is double counted by the device indicated by the red arrow.
		}
		\label{fig:egms}
	\end{center}
\end{figure*}

The clinically observed physiologic intracardiac signals \gls{egm} on the atrial channel and the ventricular channel are listed in Table \ref{tab:Aegm} and \ref{tab:vegm} respectively. The sources of these signals can be found in the given references. The clinical significance column indicates how often these problems could occur in real patients. The corresponding variables in our model to control the presence of the signals are also given. While all of these signals can be reproduced by the proposed model, we only present some examples (Fig. \ref{fig:egms}) in this paper due to the space limitation. 

\begin{table}[!h]
	
	\begin{tabu}to \columnwidth {|m{2.2cm}|X|m{1.4cm}|m{0.9cm}|}
		
		\hline
		\textbf{AEGMs}  & \textbf{Clinical Significance}& \textbf{Controls}& \textbf{Segment}\\
		\hline
		P-wave & The desired signal &  $ a,b,c_{va} $;& $ A$\\
		\hline
		P-wave double counting \cite{Burri20171031,almehairi2014unusual} & Rare \cite{Burri20171031} & Atrial conduction &$ D $ \\
		\hline
		Far-field R-wave (FFRW)&Relatively frequent\cite{Burri20171031}  & $ a,b $ & $ B $\\
		\hline
		VA crosstalk \cite{Burri20171031}& Rare \cite{Burri20171031} & $ c_{va} $ &-- \\
		\hline
	\end{tabu}
	
	\caption{Intracardiac signals of atrial EGMs }	
	\label{tab:Aegm}	
\end{table}

\begin{table}[!h]
	
	\begin{tabu}to \columnwidth {|m{2.0cm}|X|m{1.3cm}|m{0.9cm}|}
		
		\hline
		\textbf{VEGMs}  & \textbf{Clinical Significance}& \textbf{Controls}& \textbf{Segment}\\
		\hline
		R-wave  & The desired signal&  $d,b,e, c_{av} $ &$ A $\\
		\hline
		Far-field P-wave \cite{Burri20171031} & Rare in adults \cite{Swerdlow2017114} for \gls{icd} and uncommon for pacemaker \cite{Burri20171031} &  $d$ & --\\
		\hline
		R-wave double counting \cite{Swerdlow2017114}& Exceptional with pacemakers
		but may occur with ICDs \cite{Burri20171031}  & Ventricular conduction &--\\
		\hline
		T-Wave Oversensing \cite{Swerdlow2017114}  & Rare in pacemakers and more frequent in ICDs \cite{Burri20171031} & $e $ &--\\
		\hline
		AV crosstalk \cite{Swerdlow2017114}& Rarely encountered but may still  occur \cite{Burri20171031}  & $ c_{av} $ & $ C $\\
		\hline
	\end{tabu}
	
	\caption{Intracardiac signals of ventricular EGMs}
	\label{tab:vegm}		
\end{table}

We use the heart model to produce various rhythms and present the electrical activities via the proposed IEGM module to the device. The presence of signals is varied over time controlled by the sensing controller. The simulation plots in Fig. \ref{fig:egms} show (from top to bottom): AEGM, $ AS_{raw} $ (Level High shows that AEGM signal is greater than the sensing threshold and could be detected by the device), VEGM, $ VS_{raw} $, the device sensed events $ AS,VS $ and the pacing pulses $ AP,VP $. 

Segment $ A $ in Fig. \ref{fig:egms} shows the desired AEGM and VEGM. These signals only reflect the local activations and are detected by the device once at a time.  A far-field R wave is present in Segment $ B $. The signal falls into the refractory period so no extra $ AS $ is detected by the device. In Segment $ C $, the ventricular lead senses the afterpotential of atrial pacing pulse $ AP $. We mimic the interatrial block in Segment $ D $ by slowing the conduction velocity of partial atrium such that the AEGM is prolonged. In this example, the device double counted the atrial signals.

While the dipole theory has been applied in numerous cardiac electrophysiology studies \cite{lines2003mathematical}, our integration of the classical dipole model into an abstract \gls{ha}-based heart model is novel. The simulation result shows that it is a feasible and effective approach. This integration has unique advantages:

\begin{enumerate}
	\item
	The \gls{ha} formalism is amenable to formal analysis and real-time implementations. This is vital for the verification of safety-critical cardiac devices.
	
	\item
	\gls{ha}-based models favor compositional design. This benefits future expansion of the model.
	
	\item
	
	Due to the nature of the dipole theory, the morphology of \gls{egm} can capture realistic signals by refining the heart model with 3D anatomical data. This would be necessary for implantable cardioverter defibrillator (ICD) validation.	
	\item
	
	The electrocardiogram (ECG) could also be modeled by extending the heart model with an appropriate torso model. The ECG can be used to validate the heart model.	
	
	\item
	
	Stochastic features could also be captured by extending the sensing controller with probabilistic automaton. This enables us more flexibility in the context of device validation. 
	
\end{enumerate}

As the morphology of \gls{egm} created by dipole models relies on the geometry of the heart, the current \gls{egm} is an approximation of real signals because of the abstract nature of the existing heart model. In particular, the T wave records the ventricular repolarization but cannot reflect patterns of epicardium to endocardium  activation.  For a pacemaker, the major algorithms depend on the timing of the signals rather than morphology. Therefore, the current model can meet the need of the pacemaker validation. We connected the virtual heart and the proposed IEGM module to a pacemaker model, and Fig. \ref{fig:egms} shows the device can sense the \gls{egm} and deliver pacing pulses if bradycardia is present. As we discussed, the morphology can be improved by refining the heart model for ICD validation.


\section{Conclusion}
\label{sec:conclusion}

We have developed an intracardiac electrogram (IEGM) model capturing local excitation, far-field signals and pacing afterpotentials. The results show that we are able to cover sensing problems discussed in the literature. We have integrated classical dipole model into the \gls{ha}-based virtual heart to enable the amenability to formal analysis. This is desirable for the verification of safety-critical systems. The compositional feature of \gls{ha} and the nature of the dipole theory provide us with a promising potential for future expansion of the work.


	\bibliographystyle{IEEEtran}
	
	\bibliography{IEEEabrv,Heart}

\end{document}